# Assessing Hypothetical Gravity Control Propulsion

Marc G. Millis*

*NASA Glenn Research Center, 21000 Brookpark Rd, Cleveland OH 44135*

**Gauging the benefits of hypothetical gravity control propulsion is difficult, but addressable. The major challenge is that such breakthroughs are still only notional concepts rather than being specific methods from which performance can be rigorously quantified. A recent assessment by Tajmar and Bertolami used the rocket equation to correct naive misconceptions, but a more fundamental analysis requires the use of energy as the basis for comparison. The energy of a rocket is compared to an idealized space drive for the following cases: Earth-to-orbit, interstellar transit, and levitation. The space drive uses 3.6 times less energy for Earth to orbit. For deep space travel, space drive energy scales as the square of delta-v, while rocket energy scales exponentially. This has the effect of rendering a space drive 150-orders-of-magnitude better than a 17,000-sec Specific Impulse rocket for sending a modest 5000 kg probe to traverse 5 light-years in 50 years. Indefinite levitation, which is impossible for a rocket, could conceivably require 62 MJ/kg for a space drive. Assumption sensitivities and further analysis options are listed to guide further inquires.**

## Nomenclature

$\delta$ = percent of hypothetical modification, with the following distinguishing subscripts:

$gf$ = gravitational field

$gm$ = gravitational mass

$im$ = inertia mass

$m$ = inertia mass of rocket

$p$ = inertia mass of expelled propellant

$dm$ = incremental change of mass (kg)

---

| | | |
|---|---|---|
| *dr* | = | incremental change in radius (m) |
| *dt* | = | incremental change of time (s) |
| *dv* | = | incremental change in velocity (m/s) |
| Δ*v* | = | total change in velocity (m/s) |
| *e* | = | base natural log = 2.71828… |
| *E* | = | energy (Joules) |
| *F* | = | force, thrust (Newton) |
| *G* | = | gravitational constant = $6.67 \times 10^{-11}$ m$^3$•kg$^{-1}$•s$^{-2}$ |
| *g* | = | gravitational acceleration at Earth's surface = 9.81 m•s$^{-2}$ |
| $I_{sp}$ | = | specific impulse (s) |
| *K* | = | kinetic energy (J) |
| *m* | = | mass of vehicle (kg) |
| $m_p$ | = | mass of propellant (kg) |
| $M_E$ | = | mass of the Earth = $5.98 \times 10^{24}$ kg |
| *r* | = | radius from the center of the Earth (m) |
| $r_O$ | = | radius of low Earth orbit = $6.67 \times 10^6$ m |
| $r_E$ | = | radius of Earth's surface = $6.37 \times 10^6$ m |
| *t* | = | time (s) |
| *U* | = | potential energy (J) |

**Introduction**

Estimating the potential benefits of *gravity control propulsion* is challenging because such breakthroughs are still only notional concepts (hence the term *hypothetical*). A recent publication [1] took a first step toward assessing the potential benefits, specifically using a modified rocket equation to demonstrate that naive modifications of gravity or inertia do not produce much benefit. Although this is an important first step to help correct misconceptions, it is only a first step. Here, additional assessments are offered along with discussion on their limits.

The term, *hypothetical gravity control propulsion*, can represent a whole class of idealized propulsion where the fundamental properties of gravity, inertia, or spacetime are assumed to be able to be manipulated to propel vehicles. This includes notions such as space drives,[2] warp drives,[3] wormholes,[4] inertia modification,[5,6] vacuum energy propulsion,[7] hyperspace propulsion,[8] and many others. The two major performance enhancements sought are to eliminate the need for propellant and to achieve hyper-fast travel.

It should be emphasized that none of the existing concepts for these breakthroughs have reached the relative maturity of Technology Readiness Level (TRL) 1: "Basic Principles Observed and Reported."  (TRL is a standard scale for aerospace technology maturity.[9]  Although some pertinent effects have been reported, none have been independently confirmed that lead directly to a propulsive effect.  This topic is still a fledgling inquiry, where theories and phenomena have only recently begun to be rigorously studied.  A short summary of relevant research was recently published, indicating that about a quarter of the approaches have clear opportunities for continued research, a third were found not to be viable, and the rest remain unresolved.[10]

Little has been published toward quantifying potential benefits and performance estimates can vary considerably depending on the methods and assumptions.  To pave the way for a more complete suite of assessments, the performance of a hypothetical space drive is compared to a rocket using a variety of methods.  A space drive is defined as: "an idealized form of propulsion where the fundamental properties of matter and spacetime are used to create propulsive forces anywhere in space without having to carry and expel a reaction mass".[2]  For this exercise a space drive can simply be thought as a device that converts potential energy into kinetic energy.  Since the issue of momentum conservation is addressed in the cited reference, it will not be elaborated on here.

In the analyses that follow, a key feature is that *energy* is the basis of comparison, rather than using the metrics of rocketry.  Discussion on the pitfalls of using rocket equations to assess breakthrough spaceflight follows next.

**Rocketry Metrics Inadequate for Breakthroughs**

The historic tendency when trying to gauge the value of an emerging technology is to use the metrics of the incumbent technology.  Such provisional assessments can be seriously misleading, however, when the emerging technology uses fundamentally different operating methods.  For example, the value of steamships is misleading when judged in terms of sail area and rigging (Analogy from Ref 11).  Although reduced sails and rigging are indeed a consequence of steamships, the true benefit is that shipping can continue regardless of the wind conditions and with far more maneuvering control.  Similarly, the benefits of a breakthrough in inertial or gravity control would likely surpass the operational conventions of rocketry.   Issues such as optimizing specific impulse or propellant density become meaningless if there is no longer any propellant.  Three examples are offered next to illustrate the pitfalls of using the metrics of rocketry to estimate the benefits of a space drive.

The first and common misleading practice when describing a hypothetical space drive is to view it as a rocket with an *infinite* specific impulse.  This seems reasonable at first since a higher specific impulse leads to less propellant, so an infinite specific impulse should lead to zero propellant.  As shown from **equation (1)**, however, specific impulse is a measure of the thrust ($F$) per propellant weight flow rate ($g\, dm/dt$).[12]  For a true space drive, the $dm/dt$ term would be meaningless, rendering the entire equation inappropriate for assessing propellantless propulsion.

$$I_{sp} = \frac{F}{g\frac{dm}{dt}} \quad (1)$$

Furthermore, as shown from **equation (2)**, an infinite specific impulse, $I_{sp}$, implies that a propellantless space drive would require infinite energy (substituting $I_{sp}=\infty$). Conversely, this same equation can be used to conclude that a propellantless space drive would require zero energy if there was no propellant (substituting $m_p=0$). Neither of these extremes are necessarily the case. This equation is the based on the energy imparted to the propellant from the rocket's frame of reference.[13]

$$E = \frac{1}{2}m_p(I_{sp}g)^2 \quad (2)$$

A second misleading use of the rocket equation is when contemplating the utility of inertial manipulations. If such a breakthrough were ever achieved, the implications and applications would likely extend beyond rocketry. Even if used on a rocket, there are a number of different ways to envision applying such an effect, each yielding considerably different conclusions; (1) apply it to the whole rocket system, (2) just to the propellant, (3) just to the vehicle, or (4) just to the expelled propellant. To illustrate these differences, only two of these options will be compared; the case where the entire rocket system's inertia is modified, and where only the expelled propellant is modified.

It should be remarked that there are, at present, no confirmed techniques to affect such a change in inertia. It is important to stress that this is only a hypothetical example to illustrate the sensitivity of the findings to the methods, rather than to suggest that this is a realistic potential breakthrough. Numerous variations on this analysis are possible.

[INSERT HERE]

Fig. 1 Rocket in Field-Free Space

Consider a rocket in field-free space (**Figure 1**). To derive the rocket equation, one can start with conservation of momentum, where the rocket expels an increment of propellant, *dm*, to produce an incremental change in the rocket's velocity, *dv*.

The standard equation to represent this conservation of momentum has been slightly modified into **equation (3)**, where coefficients have been inserted to represent hypothetical manipulations of the inertia of the expelled propellant, $\delta_p$, and the rocket, $\delta_m$. Values of $\delta$ greater than one imply an increase, less than one imply a decrease and a $\delta$ equal to one represents no change.

$$-v_e(\delta_p)dm = dv(\delta_m)(m-dm) \quad (3)$$

Proceeding with the normal steps to derive the rocket equation, it can be shown[10] that the final result for the delta-v ($\Delta v$) imparted to the rocket is represented by:

$$\Delta v = v_e \ln\left(\frac{m + m_p}{m}\right) \frac{\delta_p}{\delta_m} \qquad (4)$$

Consider now the implications of modifying the inertia of the *whole* rocket system, which implies equal changes to $\delta_p$ and $\delta_m$. In this circumstance there is no change at all in $\Delta v$. This null finding was one of the observations reported by Tajmar and Bertolami.[1] Alternatively, consider that *only* the inertia of the expelled propellant was increased as it accelerates out of the rocket, while the inertia of the vehicle remains the same. In this case the improvement in $\Delta v$ tracks proportionally to $\delta_p$. In other words a $\delta_p$ of 1.50, representing a 50% increase in the expelled propellant's inertia, would yield a 50% increase in $\Delta v$. **Table 1** summarizes how the different assumptions yield different results.

Table 1  Different ways to modify rocket inertia

| Modified Inertia: (From equation 4) | Propellant $\delta_p$ | Rocket $\delta_m$ | Net Effect |
|---|---|---|---|
| Unmodified | 1 | 1 | Baseline |
| Whole Rocket System | $\delta$ | $\delta$ | None |
| Just Expelled Propellant | $\delta$ | 1 | $\Delta v' = \delta \Delta v$ |

In addition to the ambiguity and wide span of results when using the rocket equation to predict the benefits of modifying inertia, this approach does not provoke the questions needed to further explore such conjectures. For example, the issue of energy conservation is not revealed from the prior equations. Although momentum was conserved in the prior example, *energy* conservation was not addressed. These equations do not provide the means to calculate the extra energy required to support this hypothetical change in propellant inertia. It is presumed that any benefit must come at some expense, and since energy is a fundamental currency of mechanical transactions, it is reasonable to expect that such a benefit requires energy.

There is, however, another more critical issue missing from the above analysis, an issue that was also omitted from the Tajmar and Bertolami analysis; namely the *equivalence principle*. The equivalence principle asserts that *gravitational* mass is identical to *inertial* mass. If the inertial mass is modified, then the gravitational mass would be similarly modified.

Addressing the *equivalence principle* is the third example to illustrate why the rocket equations are inadequate to explore hypothetical gravity control. Consider the benefits of placing a launch pad above a hypothetical gravity shield (**Figure 2**). A naive assumption would be that the reduced gravity would make it easier for the rocket to ascend as if being launched from a smaller planet. This idea was provoked from the gravity-shielding claim[14] that was later found not to be reproducible.[15]

[INSERT HERE]

**Fig. 2 Rocket Over Hypothetical Gravity Shield**

There is more than one way to interpret this situation. One can consider that the gravitational field, *g*, is modified or one can consider that the mass of the rocket above the device is modified. In the case of a modified rocket, one can further consider that just its gravitational mass is affected or, if the *equivalence principle* is in effect, that both its gravitational *and* inertial mass are *equally* affected. To explore these options, start with the following equation for a rocket ascending in a gravitational field:[16]

$$m\frac{dv}{dt} = -mg - v_e \frac{dm}{dt} \qquad (5)$$

The term on the left represents the mass and acceleration of the rocket, the middle term is the force of gravity, and the right-most term is the reaction force from the expulsion of an increment of propellant with an exhaust velocity, $v_e$, relative to the rocket.

To consider the hypothetical modifications, coefficients are inserted next to reflect modifications to the gravitational field, $\delta_{gf}$, and to the rocket's gravitational mass, $\delta_{gm}$, and its inertial mass, $\delta_{im}$. As before, values of $\delta$ greater than one imply an increase, less than one a decrease, and equal to one represents no change. Also, the equation is now rearranged to isolate the inertial terms from the gravitational terms:

$$-(\delta_{gm})m(\delta_{gf})g = (\delta_{im})\left[m\frac{dv}{dt} + v_e \frac{dm}{dt}\right] \qquad (6)$$

The left hand side represents the gravitational contributions while the right represents the inertial contributions. It can be shown that this equation results in the following representation for the $\Delta v$ of the rocket:

$$\Delta v = -\frac{\delta_{gm}}{\delta_{im}}\delta_{gf} g \Delta t + v_e \ln\left(\frac{m_{initial}}{m_{final}}\right) \qquad (7)$$

Δ*t* represents the increment of time during which propellant is expelled, and accordingly the two mass terms reflect the *initial* (higher) and *final* (lower) masses of the rocket (including its stored propellant) over this time interval. With the exception of the modification coefficients, this equation is identical to that of a normal rocket ascent in a gravitational field.

**Table 2** shows how the different possible interpretations of the hypothetical gravity shield might affect this situation. If it were assumed that the gravitational field, *g*, is modified, the result would be as naively expected; it would be the same as launching in a different gravitational environment. If, however, it is assumed that the device affects the mass of the rocket, there are two further possibilities. If the *equivalence principle* is in effect, then both the gravitational *and* inertial mass are equally affected, resulting in no change in the rocket's Δ*v*. If the equivalence principle is not in effect, then only the gravitational *or* inertial masses are affected, resulting in an analogous case to launching in a different gravitational environment.

Table 2  Different ways to modify rocket launch

| Modified Term: (From eq. 7) | Grav. Field $\delta_{gf}$ | Grav. Mass $\delta_{gm}$ | Inertial Mass $\delta_{im}$ | Net Effect |
|---|---|---|---|---|
| Unmodified Launch | 1 | 1 | 1 | Baseline |
| Gravity Modified | $\delta$ | 1 | 1 | $g' = \delta g$ |
| Rocket Gravitational Mass | 1 | $\delta$ | 1 | $g' = \delta g$ |
| Rocket Inertial Mass | 1 | 1 | $\delta$ | $g' = g/\delta$ |
| Rocket Inertia & Gravity | 1 | $\delta$ | $\delta$ | No Change |

As a side note, the inertial effects, when acting on *both* the propellant and the rocket, cancel each other out as with our prior case of a rocket in field free space.

As before, the energy implications of such situations are not illuminated with such approaches.

**Energy as a Basis of Comparison**

Although comparisons built on the incumbent methods might be useful for introductory purposes, a deeper understanding of the benefits and issues are better illustrated by using a more fundamental metric. When considering moving a mass from one place to another, energy is the fundamental currency. Using this metric, three situations will be compared next; deep space travel, Earth to orbit, and levitation.

**Deep Space Travel Energy**

To compare the energy requirements of a rocket and a hypothetical space drive, the following assumptions are used. To more fully understand the challenges, approaches and potential benefits of breakthrough propulsion, it would be fruitful to repeat the analysis using different assumptions:

- The space drive is interpreted to simply be a device that converts potential energy into kinetic energy.
- Both the rocket and the space drive are assumed to be 100% efficient with their energy conversions.
- The thrusting duration is assumed to be much shorter than the trip duration, which for interstellar travel is reasonable.
- For the rocket, constant exhaust velocity is assumed.
- Non-relativistic trip velocity and exhaust velocity are assumed.
- The energy requirements for a rendezvous mission are based on equal Δv's for acceleration and deceleration.

**Energy of a Rocket:** To compare a rocket to another method that does not require propellant, we need an equation for rocket energy where the propellant mass is represented in terms of the vehicle's empty mass and the Δv of the mission – variables shared by the space drive. A common way to calculate the total kinetic energy of a rocket system, including both the rocket and the propellant, is just to calculate the kinetic energy imparted to the propellant from the rocket's frame of reference where the rocket has zero velocity (hence a zero contribution to the total kinetic energy).[12,13] This is consistent with the previously stated assumptions.

$$E = \frac{1}{2} m_p (v_e)^2 \qquad (8)$$

Next, to convert this into a form where the rocket's propellant mass, $m_p$, is represented in terms of the exhaust velocity and the mission Δ$v$, we apply the following form of the rocket equation, which is a variation of the Tsiolkovski equation:

$$m\left(e^{\left(\frac{\Delta v}{v_e}\right)} - 1\right) = m_p \qquad (9)$$

Substituting this form of the rocket equation into the kinetic energy equation yields this simple approximation:

$$E = \frac{1}{2}(v_e)^2 m \left( e^{\left(\frac{\Delta v}{v_e}\right)} - 1 \right) \qquad (10)$$

**Specific Impulse Limits:** Before proceeding, a limit should be brought to attention. For these introductory exercises, the comparisons are limited to non-relativistic regimes. For rockets, this implies limiting the exhaust velocity to ≤ 10% lightspeed. The corresponding upper limit to specific impulse easily follows from the equation relating specific impulse to exhaust velocity:[12]

$$v_e = I_{sp} g \qquad (11)$$

Setting the exhaust velocity of 10% light-speed (beyond which relativistic effects must be considered), the limiting specific impulse is found to be:

$$(10\%)\left(3.0 \times 10^8 \frac{m}{s}\right) \geq I_{sp}\left(9.8 \frac{m}{s^2}\right) \Rightarrow I_{sp} \leq 3.0 \times 10^6 s \qquad (12)$$

**Energy for a Space Drive:** Since a space drive has been defined for this exercise as a device that converts potential energy into kinetic energy, the basic equation of kinetic energy is used to calculate the required energy, where the values of vehicle mass and mission $\Delta v$ are the same as with the rocket.

$$E = \frac{1}{2} m (\Delta v)^2 \qquad (13)$$

**Comparisons:** Two things are important to note regarding the energy differences between a rocket and a hypothetical space drive. First, the energy for a given $\Delta v$ scales as an *exponent* for a rocket and scales as *square* of the $\Delta v$ for a space drive. This by itself is significant, but it is important to point out that a rocket and a space drive treat additional maneuvers differently.

For a rocket it is conventional to talk in terms of increases to $\Delta v$ for additional maneuvers. For example, a rendezvous mission has twice the $\Delta v$ (accelerate & decelerate) than just a flyby (accelerate). For space drives, however, the additional maneuvers are in terms of additional *kinetic energy*. To illustrate this difference, consider a mission consisting of multiple maneuvers, *n*, each having the same incremental change in velocity, $\Delta v_i$. Notice the location of the term representing the number, $n$, of repeated maneuvers, $\Delta v_i$, in the following two equations. In the case of the space drive, additional maneuvers scale linearly, while for rockets they scale exponentially. This is a significant difference:

$$E = \frac{1}{2}(v_e)^2 m \left( e^{\left((n)\frac{\Delta v_i}{v_e}\right)} - 1 \right) \quad (14)$$

[Rocket Maneuvers]

$$E = (n)\frac{1}{2}m(\Delta v_i)^2 \quad (15)$$

[Hypothetical Space Drive Maneuvers]

**Numerical Example:** To put this into perspective, consider a representative mission of sending a 5000 kg probe over a distance of 5 light-years in a 50-year timeframe. This range is representative of the distance to our nearest neighboring star (4.3 light-years) and the 50-yr time frame is chosen as one short enough to be within the threshold of a human career span, yet long enough to be treated with non-relativistic equations. This equates to a required trip velocity of 10% lightspeed. The probe size of 5000 kg is roughly that of the Voyager probe plus the dry mass of the Centaur Upper Stage (4075 kg) that propelled it out of Earth's orbit.[17] The comparison is made for both a flyby mission and a rendezvous mission.

The results of the comparisons are listed in **Table 3**. The rocket case is calculated for two different specific impulses, one set at the upper non-relativistic limit previously described, and another set at an actual high value achieved during electric propulsion lab tests.[18]

**Table 3   Deep Space Energy Comparison**

**(5000-kg, 5-ly, 50-yr)**

| Energy in Joules | Flyby | Rendezvous |
|---|---|---|
| **Space Drive** | $2.3 \times 10^{18}$ | $4.5 \times 10^{18}$ |
| **Theoretical Rocket** $I_{sp}$ = 3,000,000 sec | $3.8 \times 10^{18}$ | $1.5 \times 10^{19}$ |
| **Actual Rocket** $I_{sp}$ = 17,200 sec | $10^{91}$ | $10^{168}$ |

Even in the case of the non-relativistic upper limit to specific impulse – an incredibly high-performance hypothetical rocket – the space drive uses a factor of 2 to 3 less energy. When compared to attainable values of specific impulse – values that are still considerably higher than those currently used in spacecraft – the benefits of a space drive are enormous. Even for just a flyby mission, the gain is 72 orders of magnitude. When considering a rendezvous mission, <u>the gain is almost 150 orders of magnitude</u>. Again, though these results are intriguing, they should only be interpreted as the magnitude of gains sought by breakthrough propulsion research. Other assessments and results are possible.

**Earth To Orbit Energy**

Consider next the case of lifting an object off the surface of the Earth and placing it into orbit. This requires energy expenditures both for the altitude change and for the speed difference between the Earth's surface and the orbital velocity. For the hypothetical space drive, this energy expenditure can be represented as:

$$E_{SpaceDrive} = \Delta U + \Delta K \qquad (16)$$

Where $\Delta U$ is the potential energy change associated with the altitude change, and $\Delta K$ is the kinetic energy change associated with different speeds at the Earth's surface and at orbit. The change in potential energy, which requires expending work to raise a mass in a gravitational field, is represented by:

$$\Delta U = \int_{r_S}^{r_O} G \frac{M_E}{r^2} m \, dr \qquad (17)$$

The change in kinetic energy requires solving for the orbital velocity and the velocity of the Earth's surface and can be shown to take this form:[10]

$$\Delta K = \frac{1}{2} m \left[ \left( G \frac{M_E}{r_O} \right) - \left( \frac{2\pi \, r_E}{24 hrs} \right)^2 \right] \qquad (18)$$

For the case of placing the shuttle orbiter ($m = 9.76 \times 10^4$ kg) into a typical low Earth orbit, $r_O$, (*altitude* = 400 km), the energy required is found to be $3.18 \times 10^{12}$ Joules.

To assess the required energy for a rocket to accomplish the same mission, the following equation is used:[13]

$$E = \left(\frac{1}{2} F I_{sp} g\right) t \qquad (19)$$

The parenthetical term is the rocket *power*, which is mentioned for two reasons: to show this additional form of the rocket equation and to introduce the idea of contemplating *power* in addition to just *energy*. While power implications are not explored here in detail, they constitute a fertile perspective for further study.

Entering the following values for the Space Shuttle System (extracted from "STS-3 Thirds Space Shuttle Mission Press Kit, March 82," Release #82-29), the total energy for delivering the Shuttle orbiter via rockets is found to be 1.14 x $10^{13}$ Joules.

Space Shuttle Main Engines:

    Quantity = 3

    Thrust ea, $F$ = 470 x $10^3$ lbs (2.1 x $10^6$ Newtons)

    Specific Impulse, $I_{sp}$ = 453 s

    Burn Duration, $t$ = 514 s

Solid Rocket Boosters:

    Quantity = 2

    Thrust ea, $F$ = 2.9 x $10^6$ lbs (12.9 x $10^6$ Newtons)

    Specific Impulse, $I_{sp}$ = 266 s

    Burn Duration, $t$ = 126 s

Orbital Maneuvering System Engines:

    Quantity = 2

    Thrust ea, $F$ = 6 x $10^3$ lbs (27 x $10^3$ Newtons)

    Specific Impulse, $I_{sp}$ = 313 s

    Burn Duration, $t$ = 200s

Comparing this rocket energy value to the hypothetical space drive energy, where the efficiency of both systems is assumed to be 100%, indicates that the <u>space drive is 3.58 times more energy efficient</u>. When compared to the benefits of interstellar space drives, however, this gain is small.

From these cursory analyses, space drives do not appear as attractive for launching spacecraft into low orbit as they do for high Δ*v* missions or missions that require many maneuvers. Again, such introductory comparisons should not be taken too literally. These assessments are provided to demonstrate that there are a variety of ways to assess the potential benefits of propulsion breakthroughs.

**Levitation Energy**

Levitation is an excellent example to illustrate how contemplating breakthrough propulsion is different from rocketry. Rockets can hover, but not for very long before they run out of propellant. For an ideal breakthrough, some form of *indefinite* levitation is desirable, but there is no preferred way to represent the energy or power to perform this feat. Since physics defines work (energy) as the product of force acting over distance, no work is performed if there is no change in altitude. Levitation means hovering with no change in altitude.

Regardless, there are a variety of ways to toy with the notion of indefinite levitation. A few of these approaches are listed in the next session. For now, only one approach is illustrated, specifically the nullification of gravitational potential.

An object in a gravitational field has the following defined value for its gravitational potential energy:

$$U = G \frac{M_E}{r} m \qquad (20)$$

Usually this definition is used to compare energy differences between two relatively short differences in height, $r$, but in our situation we are considering this potential energy in the more absolute sense. This equation can also be derived by calculating how much energy it would take to completely remove the object from the gravitational field, as if moving it to infinity. This is more analogous to nullifying the effect of gravitational energy. This is also the same amount of energy that is required to stop an object at the levitation height, $r$, if it were falling in from infinity with an initial velocity of zero.

Using this equation, it could conceivably require <u>62 mega-Joules to levitate 1-kg near the Earth's surface</u>. This is roughly twice as much as putting 1-kg into low Earth orbit. Again, these assessments are strictly for illustrative purposes rather than suggesting that such breakthroughs are achievable or if they would even take this form if achievable. Some starting point for comparisons is needed, and this is just one version.

**Possible Further Assessments**

As illustrated with these introductory examples, there are a number of different ways to assess the potential benefits of breakthrough physics propulsion. To continue with deeper inquiry, a variety of different missions and assumptions can be addressed. Some of these are explained next in the hopes of stimulating deeper inquiries.

**Deep Space Travel:** In the case of deep space transport, the energy was previously calculated assuming a constant exhaust velocity for the rocket and thrusting durations that were negligible compared to trip times. Although reasonable assumptions for interstellar flight, it would also be instructive to repeat the energy comparisons with assumptions of constant acceleration, constant thrust, constant power, and when optimized for minimum trip time. To further explore these notions, it would also be instructive to repeat all of these comparisons using the relativistic forms of the equations.

Newtonian equations are not the only way to ponder space drives. From the formalism of *general relativity*, there are a variety of transportation concepts that do not require propellant, including: a *gravitational dipole toroid* (inducing an acceleration field from frame-dragging effects),[19] *warp drives* (moving a section of spacetime faster-than-light),[3] *wormholes* (spacetime shortcuts),[4,20] and *Krasnikov tubes* (creating a faster-than-light geodesic).[21]

To explore these *general relativity* formalisms in the context of creating space drives introduces entirely different energy requirements than with the Newtonian versions explored in this paper. In the *general relativity* approach, one must supply enough energy to manipulate all of the surrounding spacetime so that your spacecraft naturally falls in the direction that you want it to go. This can be referred to as *creating a pseudo geodesic* – reshaping spacetime to induce the preferred freefall trajectory. Although such approaches require considerably more energy than the simple Newtonian concepts, they are nonetheless instructive.

**Earth to Orbit:** Another variation on the calculations already offered for ascent to orbit would be to repeat the assessments in the case of constant power. Another interesting investigation would be to asses the energy implications of the "gravity-shielded" launch pad.

**Levitation:** It has already been mentioned that the Newtonian treatment for work is *force times distance*; hence, no work is performed to keep an object at a fixed levitation height. There are other ways to contemplate the energy of levitation that look beyond this too-good-to-be-true zero energy requirement. In addition to the potential energy approach already calculated, which yielded an energy requirement of 62-MJ/kg to levitate an object at the Earth's surface; here are a variety of other ways to contemplate this situation:

1. Helicopter analogy: Calculate the energy and power required to sustain a downward flow of reaction mass to keep the helicopter at a fixed altitude.
2. Normal accelerated motion: Rather than assess levitation energy directly (where the mass sustains zero velocity in an accelerated frame), calculate the energy or power required to continuously accelerate a mass at 1g in an inertial frame. In this case, the normal force-times-distance can be used, but issues with selecting the assumed starting velocity and limits of integration arise.
3. Escape velocity: Another way to estimate levitation energy is to calculate the kinetic energy for an object that has achieved escape velocity. This approach actually results in the same 62-MJ/kg as when calculating the absolute potential energy.

4. Thermodynamic: Treat levitation analogously to keeping a system in a non-equilibrium state, where equilibrium is defined as free-fall motion in a gravitational field and the stable non-equilibrium condition is defined as levitation at a given height.

5. Gravity shielded flywheel: In contrast to the rocket ascent case already addressed, calculate the energy and power required to continuously increase the rotation rate of a flywheel, assuming that the hypothetical gravity shield is placed under a portion of a flywheel with a horizontal axis. Like the accelerated motion approach, this introduces issues with integration limits.

6. Damped oscillation: Calculating the energy of oscillation about a median hovering height, but where an energy cost is incurred for both the upward and downward excursions, where damping losses are included.

7. Impulse: Rather than use the force-times-distance form, use the "impulse" treatment of force-times-duration. Like the accelerated motion and gravity shielded flywheel approaches, this introduces issues with integration limits.

8. Geodesic: Using the perspective of geometric general relativity, calculate the energy required to create a local null geodesic at the surface of the Earth. By "local null geodesic " it is meant where the local freefall path is a stationary trajectory.

## Concluding Remarks

The potential benefits of breakthrough propulsion cannot be calculated yet with certainty, but crude introductory assessments show that the performance gains could span from a factor of 2 to a factor of $10^{150}$ in the amount of energy required to move an object from one point to another. To send the Shuttle into orbit using a space drive could conceivable require 3.58 times less energy. To send a 5000 kg probe to a point 5 light-years distant with a trip time of 50 years would require 150 orders of magnitude less energy. And since a rocket cannot levitate an object indefinitely, the improvement is infinite, but a value of 62 MJ/kg is possibly the amount of energy that would be required to levitate an object at the Earth's surface.

Although these analyses are only for introductory purposes, a deeper understanding of the challenges of discovering such breakthroughs can be approached through similar assessments using different assumptions and methods. In particular, recalculating the deep space trajectories assuming constant acceleration or constant power would be useful, as well as using different approaches to calculate the energy and power required to levitate an object indefinitely.

FIGURES

(Placed at end of this submission to avoid formatting glitches when translating to different formats)

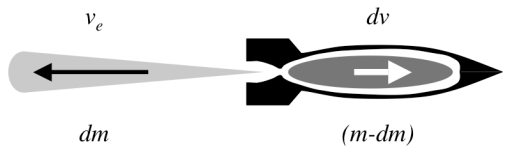

**Fig. 1  Rocket in Field-Free Space**

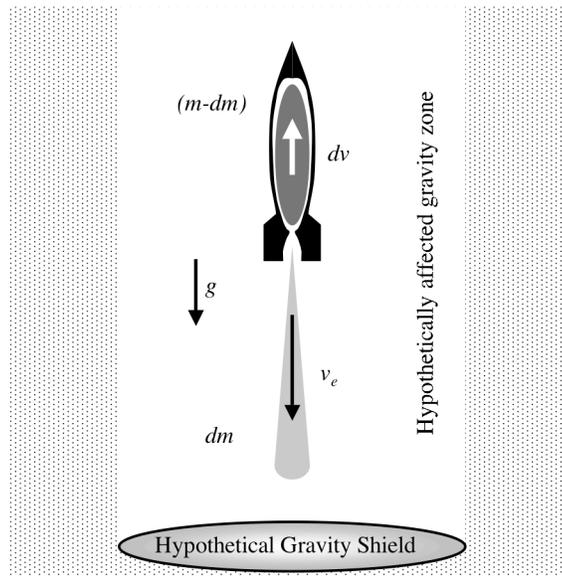

**Fig. 2  Rocket Over Hypothetical Gravity Shield**